\documentclass[a4paper,english,aps,preprint]{revtex4}
\usepackage[T1]{fontenc}
\usepackage[latin9]{inputenc}
\usepackage{amsmath}
\usepackage{graphicx}
\usepackage{amssymb}
\usepackage{esint}

\makeatletter

\newcommand{\lyxdot}{.}

\@ifundefined{textcolor}{}
{%
 \definecolor{BLACK}{gray}{0}
 \definecolor{WHITE}{gray}{1}
 \definecolor{RED}{rgb}{1,0,0}
 \definecolor{GREEN}{rgb}{0,1,0}
 \definecolor{BLUE}{rgb}{0,0,1}
 \definecolor{CYAN}{cmyk}{1,0,0,0}
 \definecolor{MAGENTA}{cmyk}{0,1,0,0}
 \definecolor{YELLOW}{cmyk}{0,0,1,0}
 }

\DeclareMathOperator{\diag}{diag}

\usepackage{ifpdf}
\ifpdf
\IfFileExists{lmodern.sty}
 {\usepackage{lmodern}}{}
 \usepackage[colorlinks=true, bookmarks, bookmarksnumbered, bookmarksopen, bookmarksopenlevel=1,
  linkcolor=black, citecolor=black, urlcolor=blue, filecolor=blue,
  pdfpagelayout=OneColumn, pdfnewwindow=true,
  pdfstartview=XYZ, plainpages=false, pdfpagelabels,
  pdfauthor={Pier Paolo Peirano, Damien Challet}, pdftex,
  pdftitle={The ups and downs of renormalization group applied to financial time series},
  pdfsubject={Quantitative Finance}]{hyperref}
\else 
 \usepackage[ps2pdf, colorlinks=true,
  linkcolor=black, citecolor=black, urlcolor=blue, filecolor=blue]{hyperref}
\fi 

\@ifundefined{showcaptionsetup}{}{%
 \PassOptionsToPackage{caption=false}{subfig}}
\usepackage{subfig}
\makeatother

\usepackage{babel}

\begin{document}

\title{The Ups and Downs of Modeling Financial Time Series with Wiener Process
Mixtures}

\author{Damien Challet}

\email{damien.challet@unifr.ch}

\affiliation{Physics Department Fribourg University\\
Pérolles, 1700 Fribourg, Switzerland}

\author{Pier Paolo Peirano}

\email{ppeirano@libero.it, corresponding author}

\affiliation{Institute for Scientific Interchange\\
Viale Settimio Severo 65, 10133 Torino, Italy}
\begin{abstract}
Starting from inhomogeneous time scaling and linear decorrelation
between successive price returns, Baldovin and Stella recently proposed
a way to build a model describing the time evolution of a financial
index. We first make it fully explicit by using Student distributions
instead of power law-truncated Lévy distributions; we also show that
the analytic tractability of the model extends to the larger class
of symmetric generalized hyperbolic distributions and provide a full
computation of their multivariate characteristic functions; more generally,
the stochastic processes arising in this framework are representable
as mixtures of Wiener processes. The Baldovin and Stella model, while
mimicking well volatility relaxation phenomena such as the Omori law,
fails to reproduce other stylized facts such as the leverage effect
or some time reversal asymmetries. We discuss how to modify the dynamics
of this process in order to reproduce real data more accurately.
\end{abstract}
\maketitle

\section{How Scaling and Efficiency Constrains Return Distribution\label{sec:bsmodel}}

Finding a faithful stochastic model of price time series is still
an open problem. Not only should it replicate in a unified way all
the empirical statistical regularities, often called stylized facts,
(cf e.g. \citet{Co01eparsfsi,BoPo03tfrdpfsptrm}), but it should also
be easy to calibrate and analytically tractable, so as to facilitate
its application to derivative pricing and financial risk assessment.
Up to now none of the proposed models has been able to meet all these
requirements despite their variety. Attempts include ARCH family (\citet{BoEnNe94am,Ts02afts-3}
and references therein), stochastic volatility (\citet{MuRu05mmfm-7}
and references therein), multifractal models (\citet{BoBoMuZu05dfmmmcb,EiKe04mmarle,BaDeMu01mftsumrw,MaFiCa97mmar}
and references therein), multi-timescale models (\citet{BoBo05msfmvf,Zu04vpvflm,ZuDaOlOl00msfm}),
Lévy processes (\citet{CoTa04fmjp-4} and references therein), and
self-similar processes (\citet{CaGeMaYo07saop}).

Recently Baldovin and Stella (B-S thereafter) proposed a new way of
addressing the question. We advise the reader to refer to the original
papers \citet{BaSt07cltasdtc,BaSt07sediem,BaSt08rssmf} for a full
description of the model as we shall only give a brief account of
its main underlying principles. Using their notation let $S(t)$ be
the value of the asset under consideration at time $t$, the logarithmic
return over the interval $[t,t+\delta t]$ is given by $r_{t,\delta t}=\ln S(t+\delta t)-\ln S(t)$;
the elementary time unit is a day, i.e., $t=0,1,\dots$ and $\delta t=1,2,\dots$days.
In order to accommodate for non-stationary features, the distribution
of $r_{t,\delta t}$ is denoted by $P_{t,\delta t}(r)$ which contains
an explicit dependence on $t$. The most impressive achievement of
B-S is to build the multivariate distribution $P_{0,1}^{(n)}(r_{0,1},\dots,r_{n,1})$
of $n$ consecutive daily returns starting from the univariate distribution
of a single day provided that the following conditions hold:
\begin{enumerate}
\item No trivial arbitrage: the returns are linearly independent, i.e. $E(r_{i,1},r_{j,1})=0$
for $i\neq j$, with the standard condition $E(r_{i,1})=0$.
\item Possibly anomalous scaling of the return distribution with respect
to the time interval $\delta t$, with exponent $D$: \[
P_{0,\delta t}(r)=\frac{1}{\delta t^{D}}P_{0,1}\left(\frac{r}{\delta t^{D}}\right)\,.\]

\item Identical form of the unconditional distributions of the daily returns
up to a possible dependence of the variance on the time $t$, i.e.
\[
P_{t,1}(r)=\frac{1}{a_{t}}P_{0,1}\left(\frac{r}{a_{t}}\right)\,.\]

\end{enumerate}
As shown in the addendum of \citet{BaSt07sediem} these conditions
admit the solution\begin{equation}
f_{0,1}^{(n)}(k_{1},\dots,k_{n})=\tilde{g}(\sqrt{a_{1}^{2D}k_{1}^{2}+\cdots+a_{n}^{2D}k_{n}^{2}}),\label{eq:bssolution}\end{equation}
where $f_{0,1}^{(n)}$ is the characteristic function of $P_{0,1}^{(n)}$,
$\tilde{g}$ the characteristic function of $P_{0,1}$, and $a_{i}^{2D}=i^{2D}-(i-1)^{2D}$.
In this way the full process is entirely determined by the choice
of the scaling exponent $D$ and the distribution $P_{0,1}$. Therefore
the characteristic function of $P_{t,\delta t}(r)$ is \[
f_{t,T}(k)=f_{0,1}^{(n)}(\underbrace{0,\dots,0}_{t\textrm{ terms}},\underbrace{k,\dots,k}_{\delta t\textrm{ terms}},0,\dots,0)=\tilde{g}(k\sqrt{(t+\delta t)^{2D}-t^{2D}}),\]
i.e.\[
P_{t,\delta t}(r)=\frac{1}{\sqrt{(t+\delta t)^{2D}-t^{2D}}}P_{0,1}\left(\frac{r}{\sqrt{(t+\delta t)^{2D}-t^{2D}}}\right).\]

The functional form of $\tilde{g}$ in Eq. \eqref{eq:bssolution}
introduces a dependence between the unconditional marginal distributions
of the daily returns by the means of a generalized multiplication
$\otimes$ in the space of characteristic functions, i.e.,\[
f_{0,1}^{(n)}(k_{1},\dots,k_{n})=\tilde{g}(a_{1}^{D}k_{1})\otimes_{\tilde{g}}\cdots\otimes_{\tilde{g}}\tilde{g}(a_{n}^{D}k_{n}),\]
 with $\otimes_{\tilde{g}}$ defined by \begin{equation}
x\otimes_{\tilde{g}}y=\tilde{g}\left(\sqrt{[\tilde{g}^{-1}(x)]^{2}+[\tilde{g}^{-1}(y)]^{2}}\right).\label{eq:bvprod}\end{equation}
At first sight this last equation may seem a trivial identity, but
it does hide a powerful statement. Suppose indeed that instead of
starting with the probability distribution $\tilde{g}$, one takes
a general distribution with finite variance $\sigma^{2}=2$ and characteristic
function $\tilde{p}_{1}$, then it is shown in \citet{BaSt07cltasdtc}
that\begin{equation}
\lim_{N\rightarrow\infty}\underbrace{\tilde{p}_{1}\left(\frac{k}{\sqrt{N}}\right)\otimes_{\tilde{g}}\cdots\otimes_{\tilde{g}}\tilde{p}_{1}\left(\frac{k}{\sqrt{N}}\right)}_{N\textrm{ terms}}=\tilde{g}(k).\label{eq:clt}\end{equation}
This means that in this framework the return distribution at large
scales is independent of the distribution of the returns at microscopic
scales: it is completely determined by the correlation introduced
by the multiplication $\otimes_{\tilde{g}}$, with fixed point $\tilde{g}$.
Note that if $\tilde{g}$ is the characteristic function of the Gaussian
distribution, then $\otimes_{\tilde{g}}$ reduces to the standard
multiplication and one recovers the standard Central Theorem Limit.

As the volatility of the model shrinks in an inexorable way, Baldovin
and Stella propose to restart the whole shrinking process after a
critical time $\tau_{c}$ long enough for the volatility autocorrelation
to fall to the noise level. In this way one recovers a sort of stationary
time series when their length is much greater than $\tau_{c}$. In
this case one expects that the empirical distribution of the return
$\bar{P}_{\delta t}(r)$ over a time horizon $\delta t\ll\tau_{c}$,
evaluated with a sliding window satisfies\begin{equation}
\bar{P}_{\delta t}(r)=\frac{1}{\tau_{c}}\sum_{t=0}^{\tau_{c}-1}P_{t,\delta t}(r).\label{eq:mixture}\end{equation}
In the original papers no market mechanism is proposed for modeling
the restart of the process; it is simply stated that the length of
different runs and the starting points of the processes could be stochastic
variables. In their simulations the length of the processes was fixed
to $\tau=500$, which corresponds to slightly more than two years
of daily data.

\section{A Fully Explicit Theory with Student Distributions\label{sec:explicit_model}}

\begin{figure}
\includegraphics[scale=0.5]{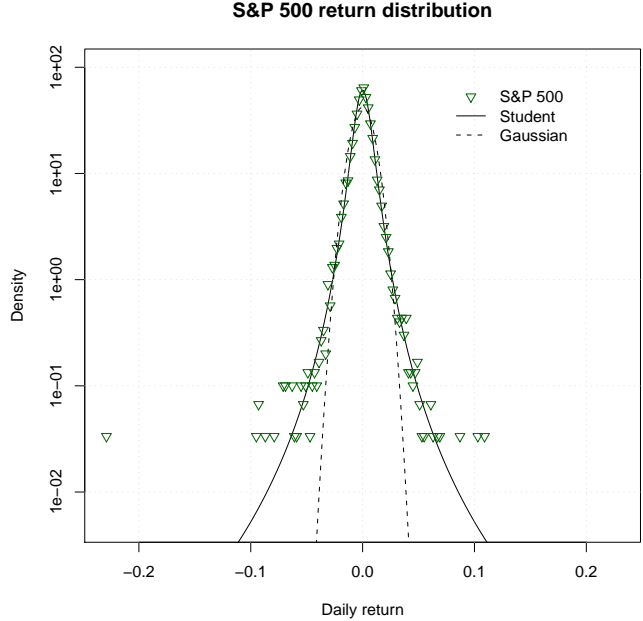}\caption{Centered distribution of the 14956 daily returns of the S\&P 500 index
(January, 3th 1950 - June, 11th 2009), and the corresponding fitting
with Student ($\nu=3.21$, $\lambda=0.0109$) and Gaussian distribution
($\sigma=0.0095$).}
\label{fig:sp500}
\end{figure}

In \citet{BaSt07sediem} a power law truncated Lévy distribution is
chosen to describe the returns\begin{equation}
\tilde{g}(k)=\exp\left(\frac{-Bk^{2}}{1+C_{\alpha}k^{2-\alpha}}\right).\label{eq:levytruncated}\end{equation}
In \citet{SoChKl04fdeplp} it is shown that this expression is indeed
the characteristic function of a probability density with power law
tails whose exponent is exponent $5-\alpha$. However, this choice
is problematic in two respects: its inverse Fourier cannot be computed
explicitly, which prevents a fully explicit theory. In addition, for
Eq.~(\ref{eq:bssolution}) to be consistent, $\tilde{g}(\sqrt{k_{1}^{2}+\cdots+k_{n}^{2}})$
must be the characteristic function of a multivariate probability
density for all $n$. In \citet{BaSt07sediem} only numerical checks
are performed to verify this property. But as discussed for example
in \citet{BoPo03tfrdpfsptrm} both truncated Lévy and Student distributions
yield acceptable fits of the returns on medium and small time scales.
In the present context, the Student distribution, sometimes referred
to as $q$-Gaussian in the case of non-integer degrees of freedom,
is a better choice; it provides analytic tractability while fitting
equally well real stock market prices (see also\citet{OsBoTs04dhsmo}).
The fit of the daily returns of the S\&P 500 index in the period with
a Student distribution

\[
g_{1}(x)=\frac{\Gamma(\frac{\nu}{2}+\frac{1}{2})}{\pi^{1/2}\lambda\Gamma(\frac{\nu}{2})}\frac{1}{(1+\frac{x^{2}}{\lambda^{2}})^{\frac{\nu}{2}+\frac{1}{2}}}\]
is reported in Fig.~\ref{fig:sp500}%
\footnote{All the graphics and numerical calculations have been performed with
\citet{Te08rlesc}.%
}.

The characteristic function of the Student density is\begin{equation}
\tilde{g}(k)=\frac{2^{1-\frac{\nu}{2}}}{\Gamma(\frac{\nu}{2})}k^{\frac{\nu}{2}}K_{\frac{\nu}{2}}(k),\label{eq:generalcharacter}\end{equation}
where $K_{\alpha}$ is the modified Bessel function of third kind.
As demonstrated in the appendix, the inverse Fourier transform of
$\tilde{g}(\sqrt{k_{1}^{2}+\cdots+k_{n}^{2}})$ for any integer $n$
is simply the multivariate Student distribution (see also \citet{ViPl07sirpqs}).
The general form of this distribution can be written as\begin{equation}
g_{n}^{(\nu)}(\mathbf{x},\mathbf{\Lambda})=\frac{\Gamma(\frac{\nu}{2}+\frac{n}{2})}{\pi^{n/2}(\det\mathbf{\Lambda})^{1/2}\Gamma(\frac{\nu}{2})}\frac{1}{(1+\mathbf{x}^{t}\mathbf{\Lambda}^{-1}\mathbf{x})^{\frac{\nu}{2}+\frac{n}{2}}}\,,\label{eq:generalq}\end{equation}
where $\nu>1$ is the exponent of the power law of the tails, $\mathcal{P}(r>R)\propto1/R^{\nu}$
and $\mathbf{\Lambda}$ is a positive definite symmetric matrix governing
the variance-covariance matrix $E(x_{i},x_{j})=\frac{\Lambda_{ij}}{\nu-2}$,
which does exist provided that $\nu>2$.

In passing, the same properties are shared by multivariate symmetric
generalized hyperbolic distributions introduced in finance by \citet{EbKe95hdf}
(see also \citet{BiKi01marhd}). The general case is obtained by an
affine change of variable, but for the sake of brevity let us restrict
to \[
f(\mathbf{x})=\frac{\alpha^{\frac{n}{2}}}{(2\pi)^{\frac{n}{2}}K_{\frac{\nu}{2}}(\alpha)}\frac{1}{(1+r^{2})^{\frac{\nu}{4}+\frac{n}{4}}}K_{\frac{\nu}{2}+\frac{n}{2}}(\alpha\sqrt{1+r^{2}})\]
for $\mathbf{x}\in\mathbb{R}^{n}$ and $r$ the usual euclidean norm
of $\mathbf{x}$. Student distributions are recovered in the limit
$\alpha\rightarrow0^{+}$. As shown in the appendix, its characteristic
function is given for any $n$ by\[
\tilde{f}_{n}(\mathbf{k})=\frac{K_{\frac{\nu}{2}}(\sqrt{\alpha^{2}+k^{2}})}{K_{\frac{\nu}{2}}(\alpha)}\frac{(\alpha^{2}+k^{2})^{\frac{\nu}{4}}}{\alpha^{\frac{\nu}{2}}}\]
 with $k=\sqrt{\sum_{i=1}^{n}k_{i}^{2}}$.

In the following we restrict the discussion to the Student distributions.
Hence we assume that the distribution of the return is given by Eq.~(\ref{eq:generalq})
with characteristic function given by Eq.~\eqref{eq:generalcharacter},
where $\mathbf{\Lambda}$ is a diagonal matrix \[
k=\sqrt{\mathbf{k}^{t}\mathbf{\Lambda}\mathbf{k}}=\lambda\sqrt{k_{0}^{2}+(2^{2D}-1)k_{1}^{2}+\cdots+(n^{2D}-(n-1)^{2D})k_{n-1}^{2}}\]
and $\lambda^{2}$ governs the variance of the returns on the time
scale chosen as a reference. Thanks to the fact that the diagonal
elements of $\mathbf{\Lambda}$ form a telescoping series the process
is indeed consistent for any number of discrete steps. Moreover it
can be generalized to the continuous time by setting, in the same
consistent way, \begin{multline}
\mathcal{P}(r_{0,\Delta t_{0}},r_{t_{1},\Delta t_{1}},\dots,r_{t_{n-1},\Delta t_{n-1}})\\
=g_{n}^{(\nu)}(r_{0,\Delta t_{0}},r_{t_{1},\Delta t_{1}},\dots,r_{t_{n-1},\Delta t_{n-1}},\mathbf{\Lambda}=\diag(t_{1}^{2D},t_{2}^{2D}-t_{1}^{2D},\dots,t_{n}^{2D}-t_{n-1}^{2D})),\label{eq:gendist}\end{multline}
where $t_{j}=\sum_{i=0}^{j-1}\Delta t_{i}$, $j\geq1$ and now $\mathbf{\Lambda}=\diag(t_{1}^{2D},t_{2}^{2D}-t_{1}^{2D},\dots,t_{n}^{2D}-t_{n-1}^{2D})$.
The existence of the continuum process is then guaranteed by the Kolmogorov
extension theorem. Starting from this expression a wider class of
processes can be generated by suitable transformations of the time,
i.e., by substituting the function $t_{i}\rightarrow t_{i}^{2D}$
for any monotonically increasing continuous function $t_{i}\rightarrow T(t_{i})$.
The process followed by the price $x(t)=\ln S(t)$ is a Student process
too, with same exponent $\nu$ and non diagonal matrix $\Lambda_{ij}=(-1)^{i+j}T(t_{\min(i,j)})$.

\begin{figure}
\subfloat[3D perspective.]{

\includegraphics[scale=0.5]{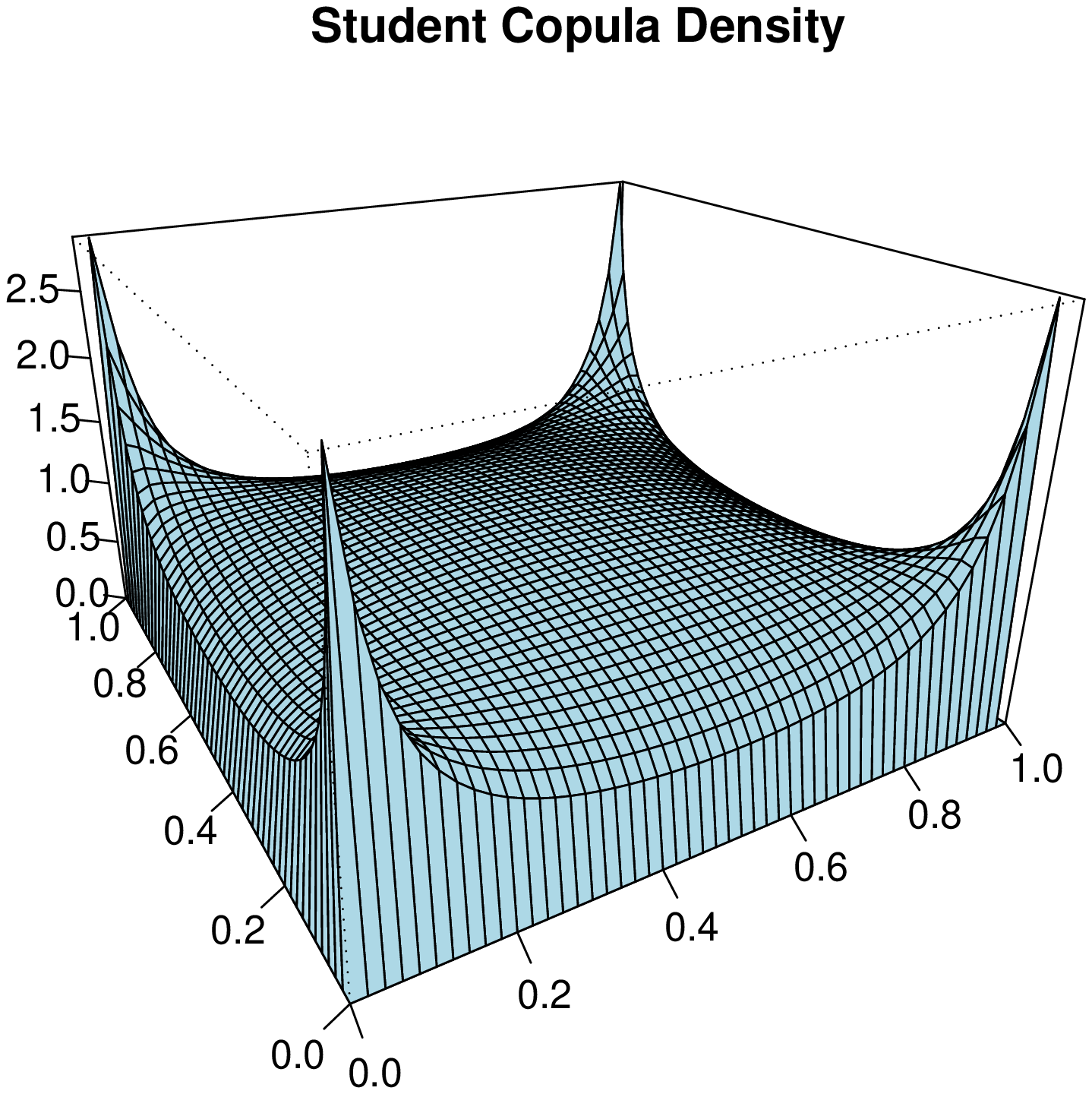}}\subfloat[Level plot.]{

\includegraphics[scale=0.5]{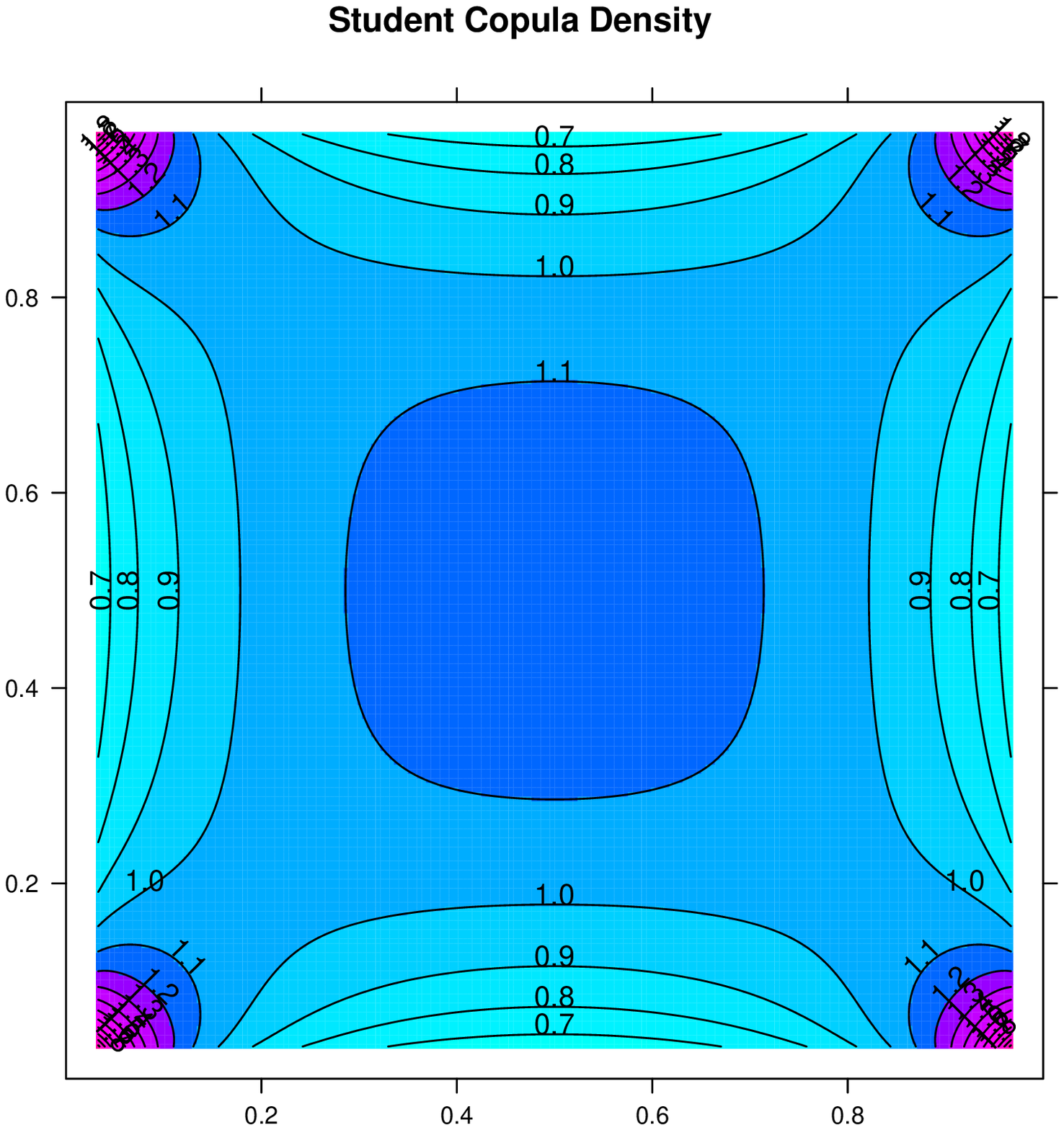}}

\caption{Student copula density with $\nu=3$ and trivial correlation matrix.}
\label{fig:copula}
\end{figure}

The Student setting makes easier to interpret the correlations induced
by the pointwise non-standard product of \eqref{eq:bvprod} in the
characteristic function space. If we consider two variables $x_{1}$
and $x_{2}$ distributed according to $g_{1}(x)$, the joint probability
function will be $g_{2}(x_{1},x_{2})$. The variables $X_{i}=G(x_{i})=\int_{-\infty}^{x_{i}}dx\, g_{1}(x)$
are distributed uniformly on the interval $[0,1]$; by definition,
the copula function $c(X_{1},X_{2})$ (cf. e.g. \citet{Ne06itc} for
a general theory) is\[
c(X_{1},X_{2})=g_{2}(G^{-1}(X_{1}),G^{-1}(X_{2}))\frac{dx_{1}}{dX_{1}}\frac{dx_{2}}{dX_{2}}=\frac{g_{2}(G^{-1}(X_{1}),G^{-1}(X_{2}))}{g(G^{-1}(X_{1}))\, g(G^{-1}(X_{2}))}.\]
In our case $c$ is none other than the Student copula function, generally
applied in finance for describing the correlation among asset prices
(\citet{ChLuVe04cmf,MaSo06efr}). A picture of this copula density
with $\nu=3$ and $\mathbf{\Lambda}$ the identity matrix is given
in Fig.\ \ref{fig:copula}. Although Student and generalized hyperbolic
distributions are usually adopted for modeling returns of several
assets over the same time intervals, the framework proposed by Baldovin
and Stella allow them to model the returns of a single asset over
different time intervals.

\section{The Baldovin-Stella Process as Multivariate Normal Variance Mixtures
\label{sec:complete_charact}}

According to the B-S framework we have to look for functions $\phi:\,\mathbb{R}\rightarrow\mathbb{C}$,
such that $\tilde{g}_{n}:\,\mathbb{R}^{n}\rightarrow\mathbb{C}$ with
$\tilde{g}_{n}(k_{1},k_{2},\dots,k_{n})=\phi(k_{1}^{2}+k_{2}^{2}+\cdots+k_{n}^{2})$
is the characteristic function of a probability distribution for any
$n$. Then from Eq.~\eqref{eq:gendist} we obtain a unique stochastic
process with a well-defined continuous limit.

B-S processes can be fully characterized if one regards their finite
dimensional marginals as instances of multivariate normal variance
mixtures $U=\sigma N$, where $\sigma$ is an univariate random variable
with positive values, $\sigma^{2}$ having cumulative distribution
$G$, and $N$ is an $n$-dimensional normal random variable independent
from $\sigma$. Leaving aside trivial affine changes of variables,
we can assume that the covariance matrix of $N$ is the identity matrix.
By first conditioning its evaluation on the value of $\sigma$, and
then computing its mean over $\sigma$, it is immediate to see that
the characteristic function $\tilde{g}_{n}^{U}(k_{1},k_{2},\dots,k_{n})$
of $U$ is\[
\tilde{g}_{n}^{U}(k_{1},k_{2},\dots,k_{n})=\phi_{\sigma^{2}}\left(\frac{1}{2}(k_{1}^{2}+k_{2}^{2}+\cdots+k_{n}^{2})\right),\]
where $\phi_{\sigma^{2}}(s)$ is the Laplace transform associated
to $G$\[
\phi_{\sigma^{2}}(s)=\int_{0}^{\infty}dx\, e^{-sx}dG(x).\]
As this construction is independent from $n$, an admissible choice
for $\phi$ is $\phi(s)=\phi_{\sigma^{2}}(\frac{s}{2})$, where $\phi_{\sigma^{2}}$
is the Laplace transform associated to any random variable $\sigma^{2}$
with positive values. 

The crucial point is that by Schoenberg's theorem in \citet{Sc42pdfos}
(see also the self-contained discussion about normal variance mixtures
in \citet{BiKi01smiftf}) this family exhausts all the possible choices,
i.e. $\phi(k_{1}^{2}+k_{2}^{2}+\cdots+k_{n}^{2})$ is a characteristic
function of a probability distribution for any $n$ if and only if
$\phi(s)$ is the Laplace transform a univariate random variable with
positive values.

Hence a multivariate distribution for the returns can be built in
the B-S framework if and only if it admits a representation as a normal
variance mixture.

In passing we note that the choice of B-S in their original papers
for the distribution \eqref{eq:levytruncated} is indeed admissible,
as in \citet{SoChKl04fdeplp} it is shown that \[
\phi_{S}(s)=\exp\left(\frac{-Bs}{1+C_{\alpha}s^{1-\alpha/2}}\right)\]
is completely monotone, hence a Laplace transform by the virtue of
Bernstein's theorem.

Now it is immediate to see that all the stochastic processes $X_{t}^{\sigma}(\omega)$
that can arise in the B-S framework admit the following representation
on a suitably chosen stochastic basis $(\Omega,\mathcal{F},\mathbb{P})$,
over which a positive random variable $\sigma(\omega)$ and a Wiener
process $W_{t}(\omega)$ independent from $\sigma$ are defined: \begin{equation}
X_{t}^{\sigma}(\omega)=\sigma(\omega)W_{t^{2D}}(\omega)\,.\label{eq:wienermixtures}\end{equation}
We only have to show that the finite dimensional marginal laws of
$X_{t}^{\sigma}(\omega)$ are the same as those arising from \eqref{eq:gendist}.
Indeed if we first evaluate the expectations over $W$, conditional
on $\sigma$, we will obtain a Gaussian multivariate distribution\begin{multline*}
\mathcal{P}(X_{t_{1}},X_{t_{2}},\dots,X_{t_{n}}\mid\sigma)\\
=\frac{1}{(2\pi\sigma^{2})^{\frac{n}{2}}}\exp\left[-\frac{1}{2\sigma^{2}}\left(\frac{X_{t_{1}}^{2}}{t_{1}^{2D}}+\frac{(X_{t_{2}}-X_{t_{1}})^{2}}{t_{2}^{2D}-t_{1}^{2D}}+\cdots+\frac{(X_{t_{n}}-X_{t_{n-1}})^{2}}{t_{n}^{2D}-t_{n-1}^{2D}}\right)\right];\end{multline*}
 the eventual average over $\sigma$ will then lead to the same multivariate
normal variance mixtures as in \eqref{eq:gendist}, with the appropriate
covariance matrix (just note that $\Delta t_{i}=t_{i+1}-t_{i}$, and
$r_{i,\Delta t_{i}}=X_{t_{i+1}}-X_{t_{i}}$). In particular, the processes
introduced in Sec.~\ref{sec:explicit_model} correspond to an inverse
Gamma distribution of $\sigma^{2}$ in the Student case, and a Generalized
Inverse Gaussian distribution in the hyperbolic case.

The stochastic differential equation obeyed by \eqref{eq:wienermixtures}
is\[
dX_{t}^{\sigma}(\omega)=\sigma(\omega)t^{D-\frac{1}{2}}dW_{t}\,,\]
This equation shows that the volatility of the processes admissible
in the B-S framework has a deterministic time dynamic, and that its
source of randomness is just ascribable to its initial value.

Eventually we can conclude that a stochastic process is compatible
with the B-S framework if and only if it is a variance mixture of
Wiener processes whose variance is distributed according an arbitrary
positive law, with a deterministic power law time change. This explains
why using use this framework to model real price returns, one inevitably
has to assume that the real price dynamics is composed by sequences
of different realizations, as done by B-S. This is necessary not only
because otherwise the model would predict a persistent and deterministic
volatility decay for $D<1/2$, but also because $\sigma$ is fixed
in each realization. The limitations of this kind of models in describing
real returns will be made more manifest in the following section,
but now we already know their mathematical foundations.

The asset prices can be modeled in an obvious arbitrage free way\[
S(t,\omega)=S_{0}\exp\left(rt+\sigma(\omega)W_{t^{2D}}(\omega)-\frac{1}{2}\sigma^{2}(\omega)t^{2D}\right)\,,\]
with $r$ the fixed default free interest rate, and where we left
the dependence on $\omega$ explicit in order to emphasise the fact
that $\sigma$ is a random variable. The pricing of options is then
the same as in the Black-Scholes model, with an additional average
over $\sigma(\omega)$. For instance the price $C(T,K)$ of a call
option with maturity $T$ and strike $K$ is\[
C(T,K)=S_{0}E_{\sigma}(N(d_{1}))-e^{-rT}KE_{\sigma}(N(d_{2}))\,,\]
with as usual $N$ is the normal cumulative distribution, \begin{alignat*}{1}
d_{1}= & \frac{\ln\frac{S_{0}}{K}+rt+\frac{1}{2}\sigma^{2}t^{2D}}{\sigma t^{D}}\,,\\
d_{2}= & \frac{\ln\frac{S_{0}}{K}+rt-\frac{1}{2}\sigma^{2}t^{2D}}{\sigma t^{D}}\,,\end{alignat*}
and the additional expectation $E_{\sigma}$ has to be evaluated according
to the distribution of $\sigma.$

\section{Applicability of this Framework to Real Markets}

The axiomatic nature of the derivation of Baldovin and Stella is elegant
and powerful: its ability to build mathematically multivariate price
return distributions from a univariate distribution using only a few
reasonable assumptions is impressive. Nevertheless, as stated in the
introduction, a model of price dynamics must meet many requirements
in order to be both relevant and useful. In this section, we examine
its dynamics thoroughly.

\subsection{Volatility dynamics}

\begin{figure}
\subfloat[Three simulations, each 500 steps long.]{

\includegraphics[scale=0.5]{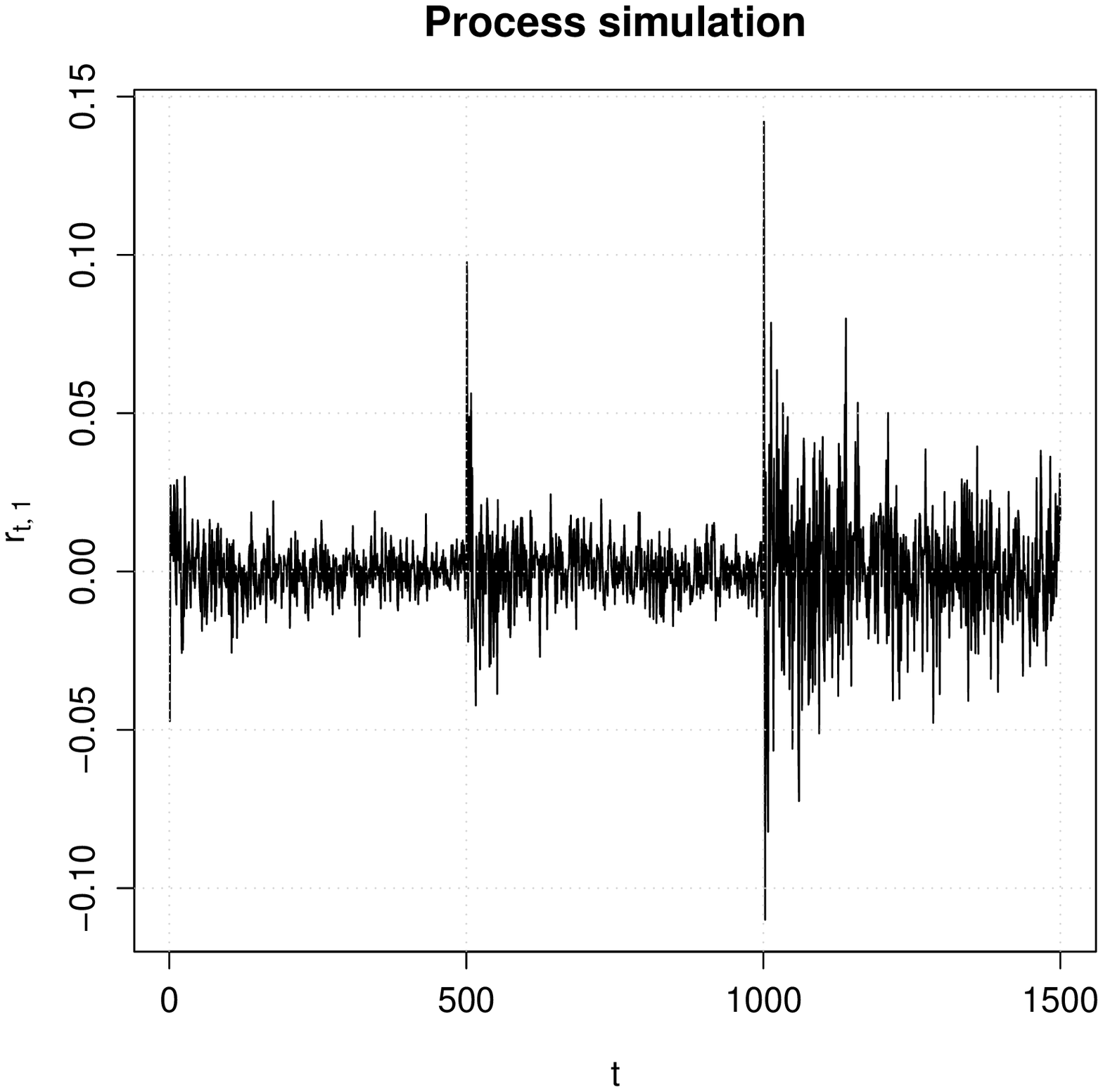}}\subfloat[Decay of the volatility: average over 10000 simulation, each 500 steps
long. The dashed line represents the analytic prediction.]{

\includegraphics[scale=0.5]{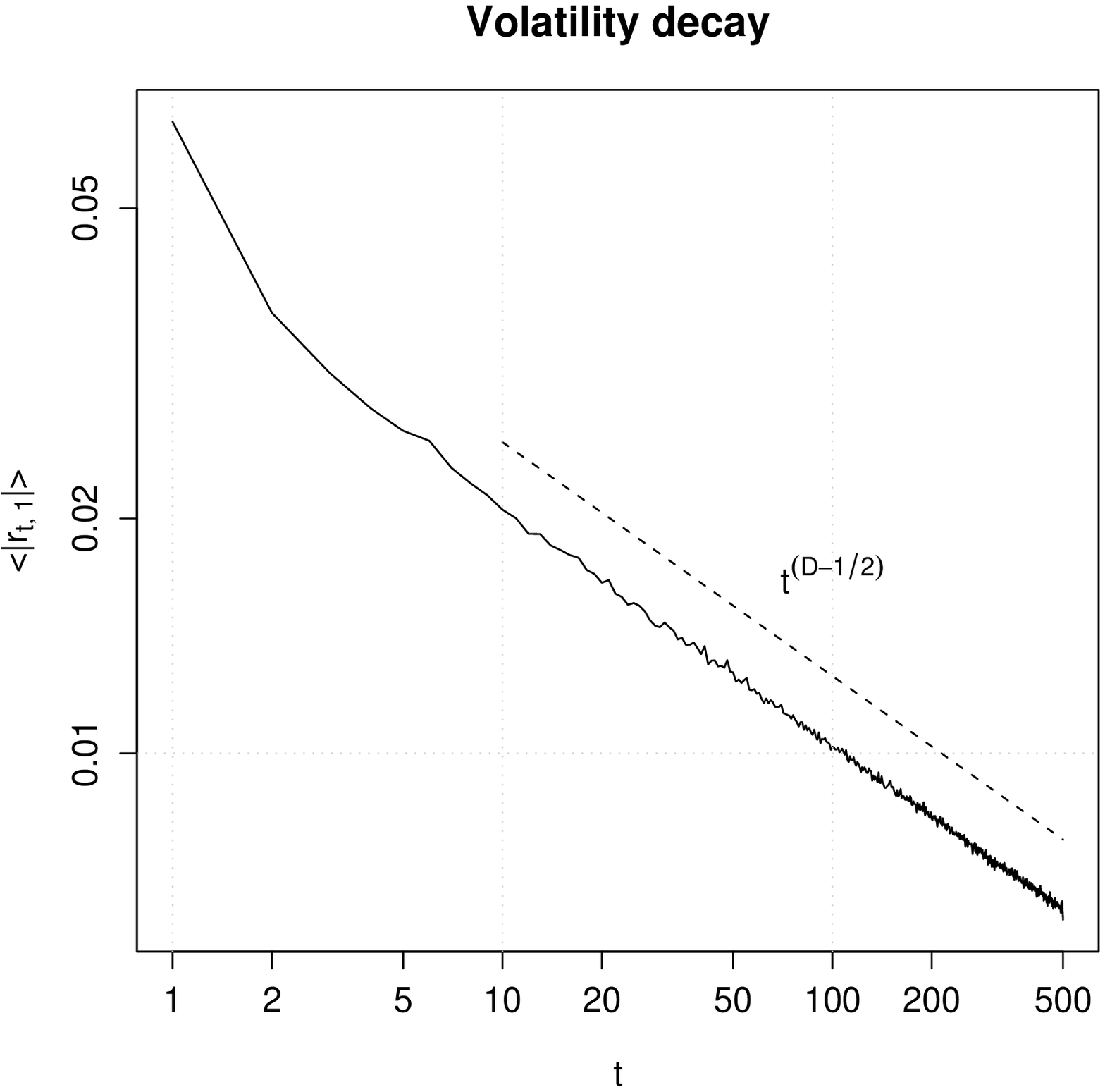}}

\caption{Process simulation with $\nu=3.2$, $D=0.20$, and $\lambda=0.107$.}
\label{fig:3runs}
\end{figure}

In Fig.~\ref{fig:3runs}.a we report the results of three simulations
of the return process, each one of 500 steps and with parameters $\nu=3.2$
and $D=0.20$. In each run the volatility decays ineluctably, as explained
in the previous section. Indeed by fixing the time interval $\delta t_{i}=1$,
we see from Eq.~\eqref{eq:gendist} that the unconditional volatility
of the $r_{t,1}$ returns is proportional to $\sqrt{(t+1)^{2D}-t^{2D}}$,
i.e., to $t^{D-1/2}$ for $t\gg1$: the unconditional volatility decreases
if $D<1/2$ and increases if $D>1/2$, in both cases according to
a power law. This appears quite clearly in Fig.~\ref{fig:3runs}.b,
where we have computed the mean volatility decay, measured as the
absolute values of the return, over 10000 process simulations. The
parameters of the distributions have been chosen close to those representing
real returns (see below).

The conditional volatility can be easily computed: the distribution
of the return $r_{n,1}$ conditioned to the previous return realizations
$r_{0,1},\dots,r_{n-1,1}$ is again a Student distribution with exponent
$\nu\prime=\nu+n$ and conditional variance\[
[(n+1)^{2D}-n^{2D}]\left(1+\sum_{i=0}^{n-1}\frac{r_{i,1}^{2}}{(i+1)^{2D}-i^{2D}}\right).\]
From this expression it is clear that volatility spikes in a given
realisation of the process tend to be persistent (see Fig.~\ref{fig:3runs}.a);
this is the main reason why fluctuation patterns differ much from
one run to an other. This can be also understood by appealing to the
characterization of this kind of processes we did in Sec.~\ref{sec:complete_charact}:
each single run is just a realization of a Wiener process, whose variance
is chosen at the beginning according to an Inverse Gamma distribution
$R\Gamma(\frac{\nu}{2},\frac{\lambda}{2})$, and that decays in time
according to the deterministic law $t^{D-\frac{1}{2}}$.

\subsection{Decreasing volatility and restarts}

The very first model introduced by B-S has constant volatility, which
corresponds to $\mathbf{\Lambda}$ being a multiple of the identity
matrix. This unfortunate feature is the main reason behind the introduction
of weights, whose effect is akin to an algebraic stretching of the
time, or, as put forward by B-S, to a time renormalization. This in
turn causes a deterministic algebraic decrease of the expectation
of the volatility, as explained above and depicted in Fig.~\ref{fig:3runs}.b;
hence the need for restarts, each attributed to an external cause.

\begin{figure}
\includegraphics[scale=0.5]{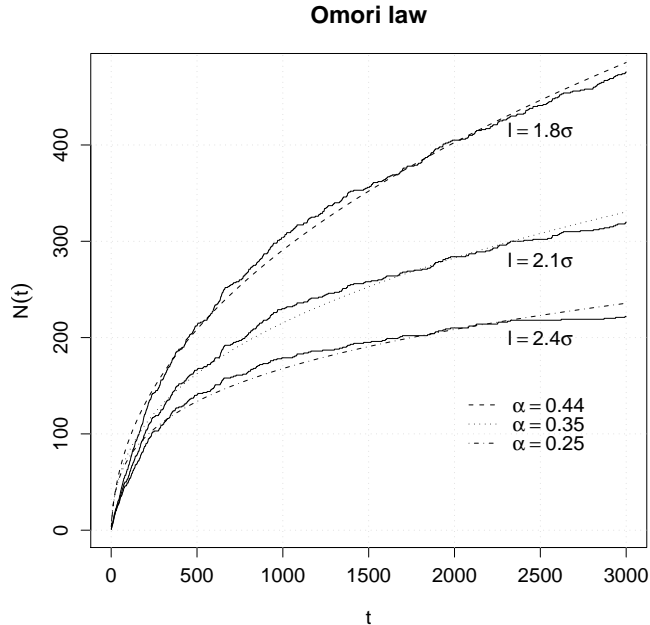}\caption{Omori law for a single run of the process, with $D=0.20$, $\nu=0.32$.
$N(t)$ is the cumulative number the absolute value of the return
exceeds a given thresholds. Three different values of the threshold
$l$ have been chosen, measured with respect to the standard deviation
$\sigma$ of the data. The dashed lines represents the fit with the
Omori law $N(t)=K(t+t_{0})^{\alpha}-Kt_{0}^{\alpha}$. }

\label{fig:omori}
\end{figure}

Although this dynamics may seem quite peculiar, such restarts are
found at market crashes, like the recent one of October 2008, which
are followed by periods of algebraically decaying volatility. This
leads to an analogous of the Omori law for earthquakes, as reported
in \citet{LiMa03prcsolafmc} and \citet{WeWaVoHaSt07rbvcfmopoas}.
The B-S model, by construction, is able to reproduce this effect in
a faithfully way. In Fig.~\ref{fig:omori} the cumulative number
of times the absolute value of the returns $N(t)$ exceeds a given
thresholds is depicted, for a single simulation of the process and
three different value of the threshold. The fit with the prediction
of the Omori law $N(t)=K(t+t_{0})^{\alpha}-Kt_{0}^{\alpha}$ is evident.

Crashes are good restart candidates: they provide clearly defined
events that synchronize all the traders' actions. In that view, they
provide an other indirect way to measure the distribution of timescales
of traders, which are thought to be power-law distributed (\citet{Li07lopaumpopldlop}).

Another example of algebraically decreasing volatility was recently
reported by \citet{McBaGu07memhssf} in foreign exchange markets in
which trading is performed around the clock. Understandably, when
a given market zone (Asia, Europe, America) opens, an increase of
activity is seen, and vice-versa. Specifically, this work fits the
decrease of activity corresponding to the afternoon trading session
in the USA with a power-law and finds an algebraic decay with exponent
$\eta=0.35$; this is exactly the same behavior as the one of B-S
model between two restarts, with $D=1-2\eta=0.3$. No explanation
of why the trading activity should result in this specific type of
decay has been put forward in our knowledge. In this case the starting
time of the volatility decay corresponds to the maximum of activity
of US markets.

\subsection{Apparent multifractality}

The Baldovin and Stella model is able to reproduce the apparent multifractal
characteristics of the real returns, i.e. the shape of $\zeta(q)$
where $\langle|r_{\delta t}|^{q}\rangle=\delta t^{\zeta(q)}$. 

\begin{figure}
\subfloat[Fitting of the empirical exponents of real data.]{

\includegraphics[scale=0.5]{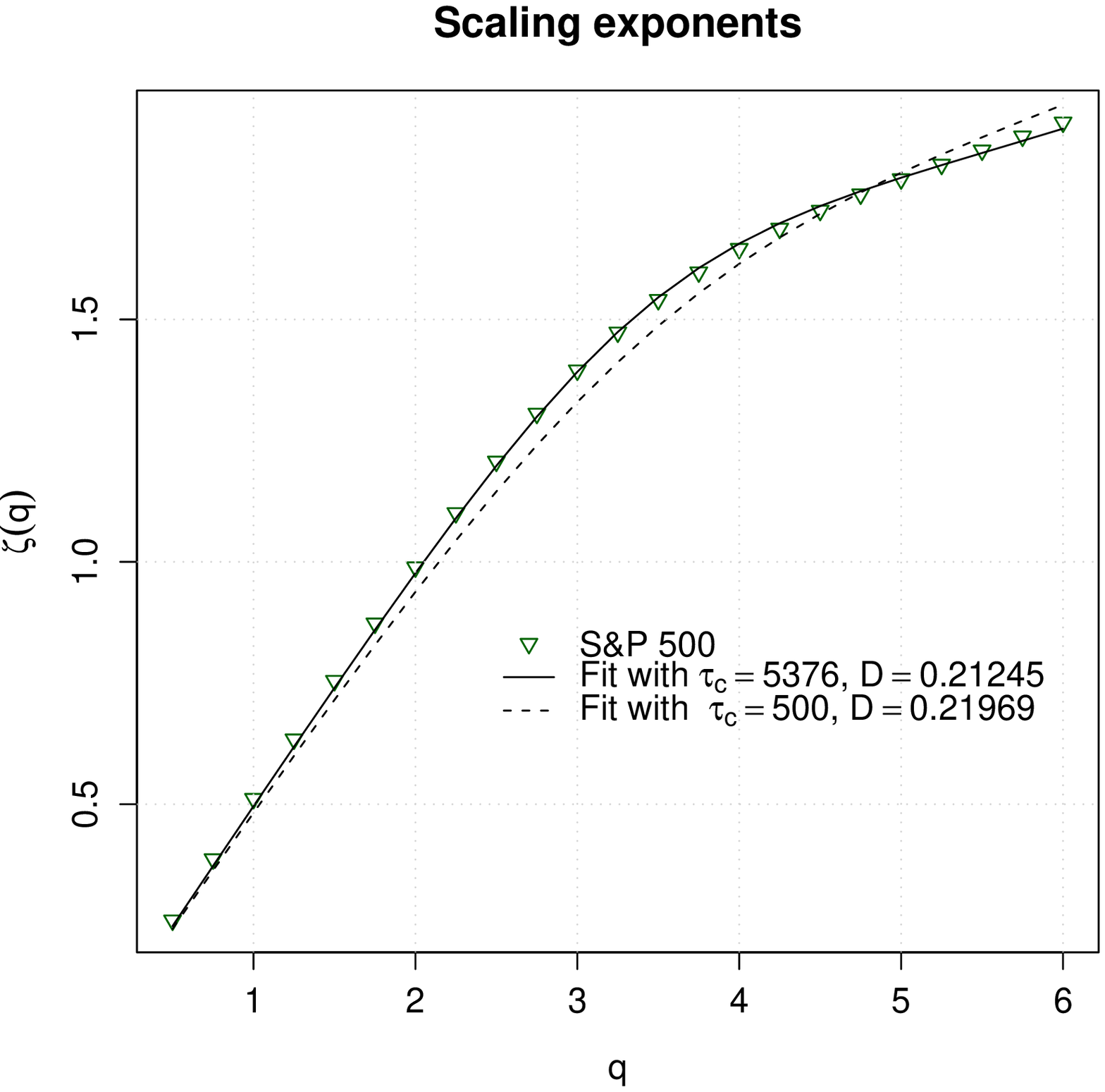}}\subfloat[Theoretical prediction compared to 5 simulations done with the same
parameters.]{

\includegraphics[scale=0.5]{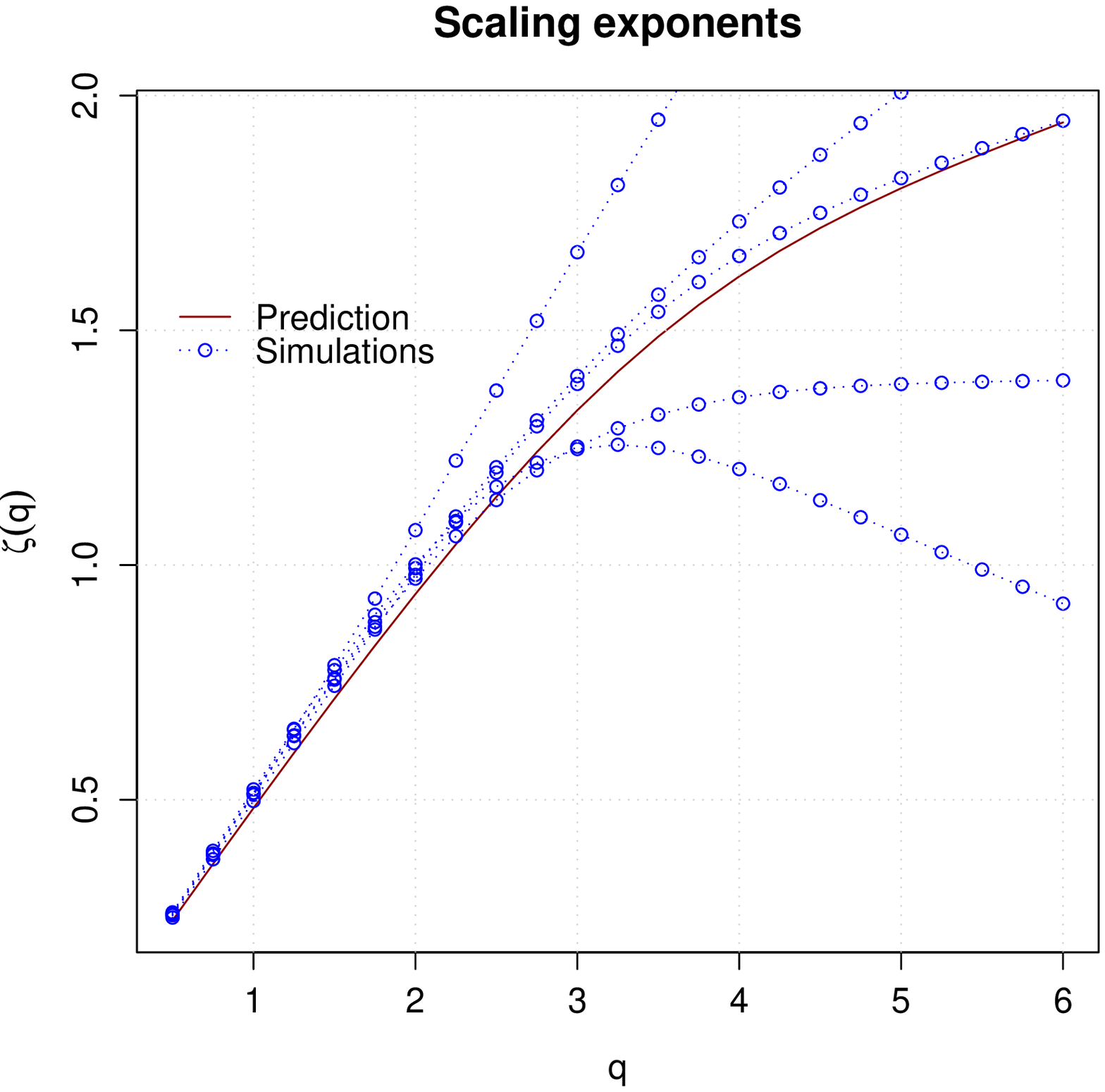}}

\caption{Scaling exponents: S\&P 500 data and simulations compared with theoretical
prediction. All the simulations have been done with the same parameters:
30 runs of 500 steps, with $\nu=3.2$, $D=0.220$}
\label{fig:scaling}
\end{figure}

The expectation is evaluated according the distribution \eqref{eq:mixture},
i.e. taking the mean over independent runs of the process. Hence the
expectation of the $q$th moment in this model is \begin{equation}
\langle|r|^{q}\rangle_{\bar{P}_{\delta t}}=\frac{\langle|r|^{q}\rangle_{P_{t=0,\delta t=1}}}{\tau_{c}}\sum_{t=0}^{\tau_{c}-1}[(t+\delta t)^{2D}-t^{2D}]^{q/2}\label{eq:scalingmodel}\end{equation}
(see the addendum to \citet{BaSt07sediem}). The exponents $\zeta(q)$
are evaluated as the slopes of the linear fitting of $\ln(\langle|r|^{q}\rangle_{\bar{P}_{\delta t}})$
with respect to $\ln(\delta t)$. Hence in our case they are determined
by the expression $\ln\sum_{t=0}^{\tau_{c}-1}[(t+\delta t)^{2D}-t^{2D}]^{q/2}$,
and depend only on $D$ and $\tau_{c}$. In Fig.~\ref{fig:scaling}.a
is depicted the fitting of the S\&P 500 exponents with the model \eqref{eq:scalingmodel}.
The best fit is obtained with $D=0.212$ and $\tau_{c}=5376$. Unfortunately
a value of $\tau_{c}$ that large is difficult to justify, as in the
case of S\&P 500 we have only 14956 daily returns, i.e. less than
three runs of a process with such a length. The other fit is obtained
by first fixing $\tau_{c}=500$, as in \citet{BaSt07sediem} and yields
$D=0.220$.

The statistical significance of this approach seems anyway questionable.
In Fig.~\ref{fig:scaling}.b we compare the theoretical expectation
of the exponents with simulations. We choose the parameters $\tau_{c}=500$,
$D=0.220$ both for simulations and analytic model, with $\nu=3.22$.
The number of restarts in the simulation is 30 in order to have a
number of data points similar to the S\&P 500. It is evident that
the exponents evaluated from the simulated data have a really large
variance.

The problem is that if the tail exponent $\nu=3.22$, from an analytic
perspective the moments with $q>3.22$ are infinite, hence, should
not be taken into account in the multifractal analysis (for an analytic
treatment of multifractal analysis see \citet{Ri02mp,Ja97mffpirvaf,Ja97mffpisf}).
The situation is somehow different in the case of multifractal models
of asset returns (\citet{BaDeMu01mrw,MaFiCa97mmar}), where the theoretical
prediction of the tail exponents of the return distribution is relatively
high (see the review of \citet{BoBoMuZu05dfmmmcb}), and the moments
usually empirically measured do exist even from the analytic point
of view. For attempts to reconcile the theoretical predictions of
the multifractal models with real data see \citet{BaKoMu06aartertvlc}
and \citet{MuBaKo06evftmf}.

It is worth remembering that the anomalous scaling of the empirical
return moments does not imply that the return series has to be described
by a multifractal model, as already pointed out some time ago in \citet{Bo99etfr}
and \citet{BoPoMe00amfts}: the long memory of the volatility is responsible
at least in part for the deviation from trivial scaling. A more detailed
analysis of real data reported in \citet{JiZh08msifof} seems indeed
to exclude evident multifractal properties of the price series.

\section{Missing Features}

Since in this model the volatility is constant in each realization
and bound to decrease unless a restart occurs, it is quite clear that
it does not contain all the richness of financial market price dynamics.
Restarting the whole process is not entirely satisfactory, as in reality
the increase of volatility is not always due to an external shock.
Volatility does often gradually build up through a feedback loop that
is absent from the B-S mechanism. Thus, large events and crashes can
also have a endogenous cause, e.g.\ due to the influence of traders
that base their decisions on previous prices or volatility, such as
technical analysts or hedgers. A quantitative description of this
kind of phenomena is attempted for instance in \citet{So03cmc,SoMaMu03vfolseve},
by appealing to discrete scale invariance (see also the viewpoint
expressed in \citet{ChFe06abaolptfc} and references therein). This
kind of effect is completely missing from the original B-S mechanism.

Volatility build-ups can be simulated with $D>1/2$, getting at constant
$D$ the equivalent of the inverse Omori law for earthquakes \citep{HeSoGr03maacfhdfspefal}.
This kind of dynamics has been reported to happen prior to some financial
market crashes \citep{SoMaMu03vfolseve}. At a smaller time scale,
foreign exchange intraday volatility patterns have a systematically
increasing part whose fit to a possibly arbitrary power-law, as performed
in \citet{McBaGu07memhssf} ($\eta=0.22$), corresponds indeed to
choosing $D=0.56$. To our knowledge, volatility build-ups either
do not follow a particular and systematic law, or perhaps have not
yet been the objects of a thorough study.

Because of the symmetric nature of all the distributions derived above,
all the odd moments are zero, hence, the skewness of real prices cannot
be reproduced. This shows up well in Fig.~3 of \citet{BaSt08rssmf}.
Another consequence is that it is impossible to replicate the \emph{leverage
effect}, i.e. the negative correlation between past returns and future
volatility, carefully analyzed in \citet{BoMaPo01lefmrvm}.

\begin{figure}
\subfloat[]{

\includegraphics[scale=0.5]{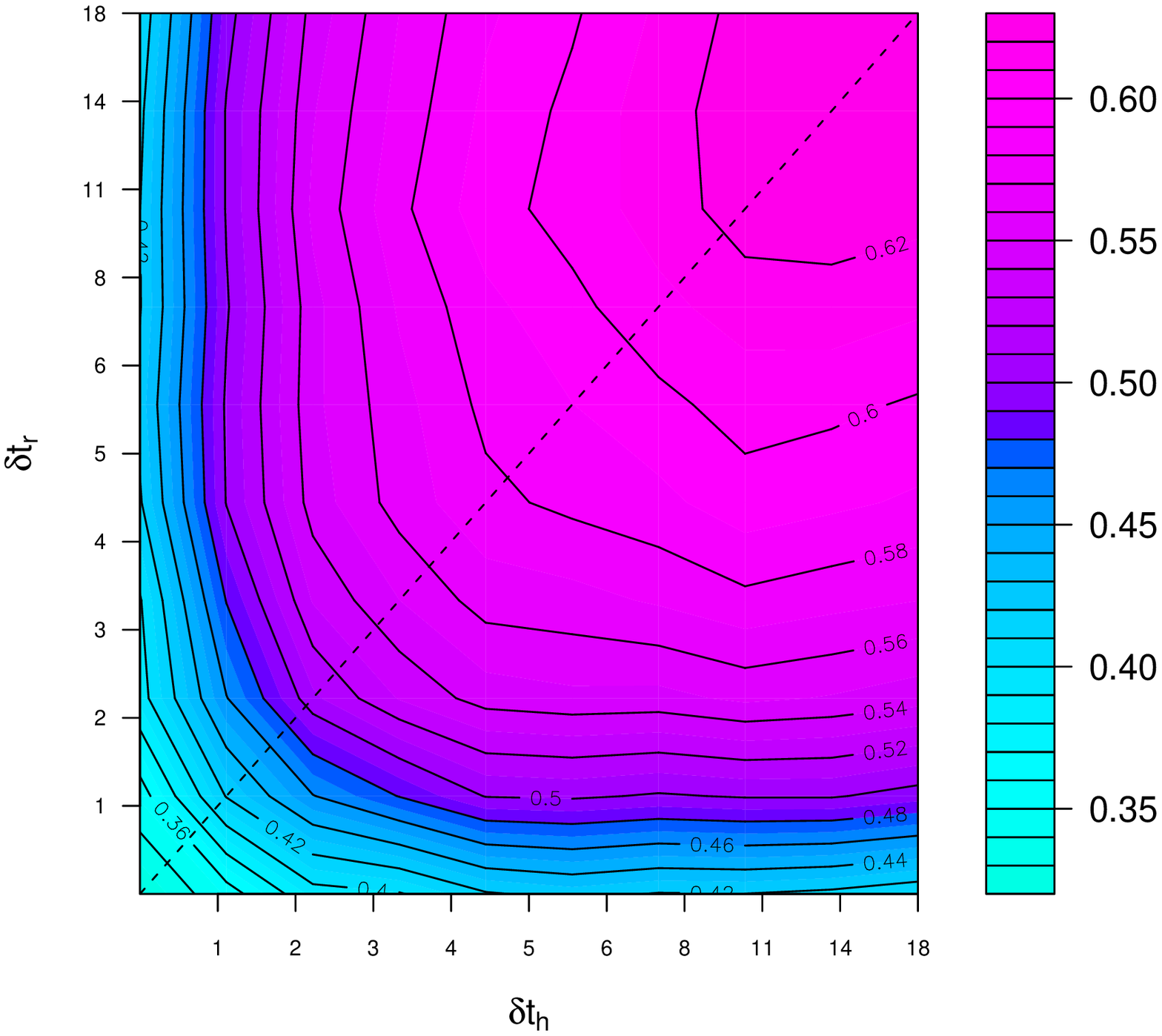}}\subfloat[]{

\includegraphics[scale=0.5]{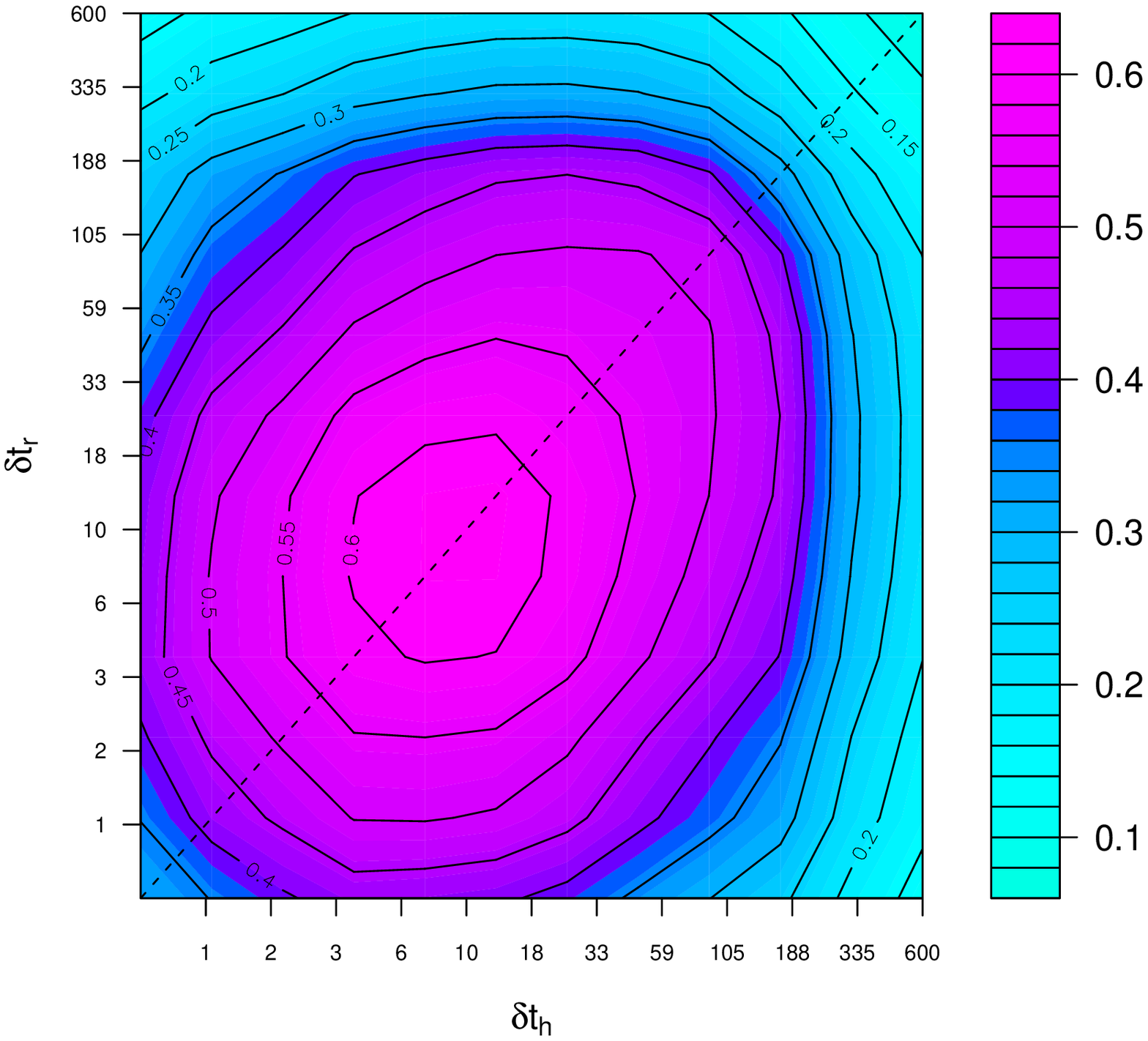}}

\caption{Correlation between historical and realized volatility of the simulated
process, over different time interval $\delta t$. The analyzed time
series was composed by 1000 runs of the basic process, each one with
200 steps, and parameter $\nu=3.22$, $D=0.20$.}
\label{fig:mugshot}
\end{figure}

In any case, the decrease of the fluctuations in the B-S process is
a deterministic outcome of the anomalous scaling law $t^{D}$ with
$D<1/2$, and results in a strong temporal asymmetry of the corresponding
time series. But quite remarkably it misses the time-reversal asymmetry
reported in \citet{LyZu03mhcsv} and \citet{Zu07trif}. Indeed real
financial time series are not symmetric under time reversal with respect
to even-order moments. For instance, there is no leverage effect in
foreign exchange rates, and their time series are not as skewed as
indices, but they do have a time arrow. One of the indicators proposed
in \citet{LyZu03mhcsv} is the correlation between historical volatility
$\sigma_{\delta t_{h}}^{(h)}(t)$ and realized volatility $\sigma_{\delta t_{r}}^{(r)}(t)$.
The historical volatility series $\sigma_{\delta t_{h}}^{(h)}(t)$
represents the volatility computed using the data in the past interval
$[t-\delta t_{h},t]$, and $\sigma_{\delta t_{r}}^{(r)}(t)$ represents
the volatility computed using the data in the future interval $[t,t+\delta t_{r}]$;
the correlation between the two series is then analyzed as a function
of both $\delta t_{r}$ and $\delta t_{h}$. Real financial time series
present an asymmetric graph with respect the change $\delta t_{h}\leftrightarrow\delta t_{s}$,
with a strong indication that historical volatility at a given time
scale $\delta t_{h}$ is more likely correlated to realized volatility
with time scale $\delta t_{r}<\delta t_{h}$, with peaks of correlation
at time scales related to human activities. The asymmetry characteristic
is absent in the Baldovin and Stella model, as showed in Fig.~\ref{fig:mugshot}.

\begin{figure}
\includegraphics[scale=0.5]{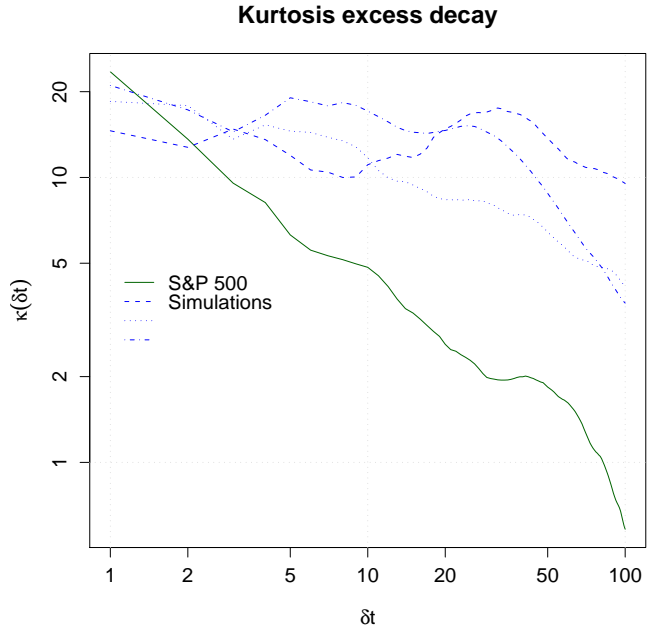}

\caption{Comparison of the kurtosis of the returns evaluated over a time interval
$\delta t$. Each one of the three simulations are composed by 30
runs, 500 steps long, in order to have a length comparable with that
of the S\&P 500 returns. The parameters are $\nu=3.2$, $D=0.20$,
$\lambda=0.1$.}
\label{fig:kurtosis}
\end{figure}

The strong correlation between returns guarantees the slow decay of
the volatility but induces some side effects. The distribution of
the returns in the model is essentially the same with identical power
law exponent for the tails. This happens independently of the time
interval $\delta t$ over which the returns are evaluated, as long
as $\delta t\ll\tau_{c}$, with $\tau_{c}$ of the order of hundreds
days. Hence the weekly returns are distributed as the daily returns,
while in real data the tail exponent begins to increase in a remarkable
way already at the intraday level (\citet{DrFoKwOsRa07smrdfptp}).
The strong correlation also slows down the convergence to the Gaussian
distribution of the returns when measured on larger time scale. Even
if the kurtosis is not defined analytically in principle, it is possible
to measure the empirical kurtosis of the returns of a simulated time
series and compare with the kurtosis of real data. In Fig.~\ref{fig:kurtosis}
we show the kurtosis of the return distribution among simulations
and daily return of the S\&P 500 index; the kurtosis has been computed
for the returns over different interval $\delta t$, and the simulated
processes had the same length (30 runs of 500 steps) of the real series.

\section{Suggested Improvements}

The main limitations of the model proposed by Baldovin and Stella
are poor volatility dynamics, lack of skewness, some unwanted symmetry
with respect to time, and extremely slow convergence to a Gaussian.
In this final section we put forward briefly some qualitative proposals
of how these issues can be addressed.

The volatility dynamics can be improved by introducing an appropriate
dynamics for the exponent $D$, i.e. introducing a dynamic $D(t)$
controlling the diffusive process. This is equivalent to starting
with a model with constant volatility, i.e. with $\mathbf{\Lambda}$
proportional to the identity matrix, and then introducing an appropriate
evolution for the time $t$. This technique is employed for instance
in the Multifractal Random Walk model (\citet{BaDeMu01mrw}), where
the time evolution is driven by a multifractal process, or when the
time evolution is modeled by an increasing Lévy process (see e.g.
\citet{CoTa04fmjp-4}). In this last case we would obtain a mixing
of Wiener processes driven by a subordinator.

The lack of skewness is a common problem of stochastic volatility
models: one usually writes the return at time $t$ as $r_{t,\delta t}=\epsilon(t)\sigma(t)$,
where $\epsilon(t)$ is sign of the return and $\sigma(t)$ its amplitude,
a symmetric setting if the distribution of $\epsilon(t)$ is even.
One remedy found for instance in \citet{EiKe04mmarle} is to bias
the sign probabilities while enforcing a zero expectation; more precisely,
\[
P\left(\epsilon=\pm\frac{1/\sqrt{2}}{1/2\pm\epsilon}\right)=1/2\pm\epsilon.\]
Another possibility for introducing skewness is that of considering
normal mean-variance mixtures, instead of simply normal variance ones.
For instance, this would have implied the use of the multivariate
skewed Student distribution in the model described in Sec.~\ref{sec:explicit_model}.

The decay of the tail exponent of the return distribution, represented
in Fig.~\ref{fig:kurtosis}, could be implemented by introducing
two different Student distributions: a univariate with exponent $\nu_{r}$
for modeling the daily returns, and a multivariate one with a much
larger exponent $\nu_{c}$ for modeling the correlations among them.
By taking into account the generalized central limit theorem expressed
in Eq.~\eqref{eq:clt}, the distribution of returns at intermediate
time scales will interpolate between the two exponents, yielding the
desired feature.

The Zumbach mugshot is one of the most difficult stylized facts to
reproduce. To our knowledge the best results in that respect was achieved
in \citet{BoBo05msfmvf}, where a specific realization of a quadratic
GARCH model is introduced, motivated by the different activity levels
of traders with different investment time horizons, which take into
account the return over a large spectrum of time scales. More specifically
Borland and Bouchaud use

\[
\sigma_{i}^{2}=\sigma_{0}^{2}\left[1+\sum_{\delta t=1}^{\infty}\mathit{g_{\Delta t}}\frac{r_{i,\delta t}^{2}}{\sigma_{0}^{2}\tau\delta t}\right],\]
with $\tau$ fixing the time scale, $r_{t,\delta T}=\ln S(t+\delta T)-\ln S(t)$,
$g_{\delta t}$ measuring the impact on the volatility by traders
with time horizon $\delta t$, and chosen by the authors $g_{\delta t}=g/(\delta t)^{\alpha}$.
This expression is rewritten also in the form\[
\sigma_{i}^{2}=\sigma_{0}^{2}+\sum_{j<i,k<i}\mathcal{M}(i,j,k)\frac{r_{j}r_{k}}{\tau},\]
with \[
\mathcal{M}(i,j,k)=\sum_{\Delta t=\max(i-j,i-k)}^{\infty}\frac{g_{\delta t}}{\delta t}.\]

In the present framework this would correspond to use a highly non-trivial
matrix $\mathbf{\Lambda}$, introducing linear correlation among returns
at any time lag. This means that the B-S process would no longer be
a model of returns, but of stochastic volatility.

\section{Discussion and Conclusions}

When employed with self-decomposable distributions like the Student
or the Generalized Hyperbolic as introduced in Sec.~\ref{sec:explicit_model},
the resulting description of the process return is different than
that of other models in the literature. First our Student process
is not stationary, hence different from the class of Student processes
discussed in \citet{HeLe05sp}, where the main focus is on stationary
ones. The processes \eqref{eq:wienermixtures} are also different
from the one studied in \citet{Bo02opfbnspm}: the latter too are
continuous and based on the Student distributions, but defined by
the stochastic differential equation

\[
dX_{t}=t^{D-\frac{1}{2}}\sqrt{\frac{2Dc_{0}}{\nu-1}}\sqrt{1+\frac{X_{t}^{2}}{c_{0}t^{2D}}}dW\,;\]
apart from the striking difference with Eq.~\eqref{eq:wienermixtures},
in \citet{VeNi07oopmitpoht} it is shown that not all the marginal
distribution laws of $X_{t}$ are of Student type.

Instead in \citet{EbKe95hdf} the Generalized Hyperbolic laws are
adopted for describing the returns at a fixed time scale; these laws
are then extended to the other time scales using the standard Lévy
process construction: in this case the distributions at the other
time scales are no more of Generalized Hyperbolic type.

The Baldovin and Stella model is also intrinsically simpler than the
ones described in \citet{BaSh01nomasotuife}, where the volatility
has a dynamic modeled by Ornstein-Uhlenbeck type processes,\[
d\sigma_{t}^{2}=-\lambda\sigma_{t}^{2}dt+dL_{t}\]
driven by an arbitrary Lévy process $L_{t}$. In this case, according
to the choice of $L_{t}$, any self-decomposable distribution (like
the Generalized Inverse Gaussian, or any of its special cases, like
the Inverse Gamma) can arise as the distribution of $\sigma_{t}^{2}$
for any $t$. But this simplification comes at a high price: while
in Barndorff-Nielsen $\sigma$ is truly dynamic, it is fixed in B-S
for any single process realization.

In addition, the models analyzed in \citet{CaGeMaYo07saop} are of
a different type, even if there are some analogies in the underlying
principles. In \citet{CaGeMaYo07saop} indeed an anomalous scaling
is introduced by considering self-similar processes, and in that framework
any self-decomposable distribution can employed for modeling returns,
but once again only at a fixed time scale, as in the standard case
of Lévy processes. The main difference is that in \citet{CaGeMaYo07saop}
the returns at different times are assumed to be totally independent,
but not identically distributed: instead Baldovin and Stella assume
that the returns are only linearly independent, but now with identical
distributions at all the time scales, up to a simple rescaling. 

In conclusion, despite its current inability to reproduce all the
needed stylized facts, the new framework proposed by Baldovin and
Stella introduces a new mechanism for modeling returns, based on a
few reasonable first principles. We therefore think that, once suitably
modified for instance along the lines proposed above, the B-S framework
can provide a new tool for building models of financial price dynamics
from reasonable assumptions.

\section*{Appendix: Some Useful Facts About Student and Symmetric Generalized
Hyperbolic Distributions}

\subsection*{Characteristic function of Student distributions}

The standard form of univariate Student distribution is\[
g_{1}(x)=\frac{\Gamma(\frac{\nu}{2}+\frac{1}{2})}{\pi^{1/2}\Gamma(\frac{\nu}{2})}\frac{1}{(1+x^{2})^{\frac{\nu}{2}+\frac{1}{2}}},\]
while the multivariate one is \[
g_{n}(\mathbf{x})=\frac{\Gamma(\frac{\nu}{2}+\frac{n}{2})}{\pi^{n/2}\Gamma(\frac{\nu}{2})}\frac{1}{(1+r^{2})^{\frac{\nu}{2}+\frac{n}{2}}}\]
with $r=\sqrt{\sum_{i=1}^{n}x_{i}^{2}}$ and $\mathcal{P}(r>R)\propto1/R^{v}$.

Using some standard relationships involving Bessel functions one can
compute analytically the corresponding characteristic function:

\begin{multline*}
\tilde{g}_{1}(k_{1})=\int_{-\infty}^{+\infty}dx_{1}\, e^{ik_{1}x_{1}}g_{1}(x_{1})\\
=\frac{2\Gamma(\frac{\nu}{2}+\frac{1}{2})}{\pi^{1/2}\Gamma(\frac{\nu}{2})}k^{\nu}\int_{0}^{+\infty}dx\,(k^{2}+x^{2})^{-\frac{\nu}{2}-\frac{1}{2}}\cos(x)=\frac{2^{1-\frac{\nu}{2}}}{\Gamma(\frac{\nu}{2})}k^{\frac{\nu}{2}}K_{\frac{\nu}{2}}(k),\end{multline*}
with $k=|k_{1}|$, $K_{\alpha}$ the modified Bessel function of third
kind, and the employ of identity 7.12.(27) of \citet{erdely2}\begin{multline*}
K_{\nu}(z)=\frac{(2z)^{\nu}}{\pi^{1/2}}\Gamma(\nu+\frac{1}{2})\int_{0}^{\infty}dt\,(t^{2}+z^{2})^{-\nu-1/2}\cos(t)\\
\Re(\nu)>-\frac{1}{2},\;\mid\arg(z)\mid<\frac{\pi}{2}.\end{multline*}
For an alternative derivation we refer to \citet{Hu95cfstd} and to
the discussion in \citet{HeLe05sp}. An alternative expression is
found in \citet{DrKo02ncft}.

For general $n$ we obtain again the same expression. Indeed

\begin{multline*}
\tilde{g}_{n}(\mathbf{k})=\int_{\mathbb{R}^{n}}d^{n}\mathbf{x}\, e^{i\mathbf{k}\cdot\mathbf{x}}g_{n}(\mathbf{x})\\
=\frac{\Gamma(\frac{\nu}{2}+\frac{n}{2})}{\pi^{n/2}\Gamma(\frac{\nu}{2})}\int d^{n-2}\Omega\int_{0}^{+\infty}dr\, r^{n-1}\int_{0}^{\pi}d\phi\,\sin^{n-2}(\phi)e^{ikr\cos\phi}(1+r^{2})^{-\frac{\nu}{2}-\frac{n}{2}}\\
=\frac{2^{n/2}\Gamma(\frac{\nu+n}{2})}{\Gamma(\frac{\nu}{2})}k^{1-n/2}\int_{0}^{+\infty}dr\, r^{n/2}(1+r^{2})^{-\frac{\nu}{2}-\frac{n}{2}}J_{n/2-1}(kr)\\
=\frac{2^{1-\frac{\nu}{2}}}{\Gamma(\frac{\nu}{2})}k^{\frac{\nu}{2}}K_{\frac{\nu}{2}}(k),\end{multline*}
with $k=\sqrt{\sum_{i=1}^{n}k_{i}^{2}}$, $d^{n-2}\Omega$ the surface
element of the sphere $S^{n-2}$, $\phi$ the angle between $\mathbf{k}$
and $\mathbf{x}$ and the employ of identities 7.12.(9)\begin{multline}
\Gamma(\nu+\frac{1}{2})J_{\nu}(z)=\frac{1}{\pi^{1/2}}(\frac{z}{2})^{\nu}\int_{0}^{\pi}d\phi\, e^{iz\cos\phi}(\sin\phi)^{2\nu}\\
\Re(\nu)>-\frac{1}{2}\,,\label{eq:bess2}\end{multline}
and 7.14.(51) of \citet{erdely2},\begin{multline*}
\int_{0}^{\infty}dt\, J_{\mu}(bt)(t^{2}+z^{2})^{-\nu}t^{\mu+1}=(\frac{b}{2})^{\nu-1}\frac{z^{1+\mu-\nu}}{\Gamma(\nu)}K_{\nu-\mu-1}(bz)\\
\Re(2\nu-\frac{1}{2})>\Re(\mu)>-1,\;\Re(z)>0.\end{multline*}

Eventually one finds \[
\tilde{g}_{n}(\mathbf{k})=\tilde{g}_{1}\left(\sqrt{k_{1}^{2}+\cdots+k_{2}^{2}}\right).\]

With the linear change of variables $\mathbf{x}\rightarrow\mathbf{C}^{-1}\mathbf{x}$,
setting $\mathbf{\Lambda}^{-1}=(\mathbf{C}^{T})^{-1}\mathbf{C}^{-1}$,
i.e. $\mathbf{\Lambda}=\mathbf{C}\mathbf{C}^{T}$, one obtains the
following generalizations:\begin{equation}
g_{n}(\mathbf{x})=\frac{\Gamma(\frac{\nu}{2}+\frac{n}{2})}{\pi^{n/2}(\det\mathbf{\Lambda})^{1/2}\Gamma(\frac{\nu}{2})}\frac{1}{(1+\mathbf{x}^{t}\mathbf{\Lambda}^{-1}\mathbf{x})^{\frac{\nu}{2}+\frac{n}{2}}}\,,\label{eq:gmqg}\end{equation}
with characteristic function \[
\tilde{g_{n}}(\mathbf{k})=\frac{2^{1-\frac{\nu}{2}}}{\Gamma(\frac{\nu}{2})}(\mathbf{k}^{t}\mathbf{\Lambda}\mathbf{k})^{\frac{\nu}{4}}K_{\frac{\nu}{2}}((\mathbf{k}^{t}\mathbf{\Lambda}\mathbf{k})^{1/2}).\]

In the univariate case $\mathbf{\Lambda}$ is substituted by the scalar
$\lambda^{2}$ and the previous expressions reduce to\begin{equation}
g_{1}(x)=\frac{\Gamma(\frac{\nu}{2}+\frac{1}{2})}{\pi^{1/2}\lambda\Gamma(\frac{\nu}{2})}\frac{1}{(1+\frac{x^{2}}{\lambda^{2}})^{\frac{\nu}{2}+\frac{1}{2}}}\label{eq:gqg}\end{equation}
and\[
\tilde{g}_{1}(k)=\frac{2^{1-\frac{\nu}{2}}}{\Gamma(\frac{\nu}{2})}(\lambda k)^{\frac{\nu}{2}}K_{\frac{\nu}{2}}(\lambda k).\]

\subsection*{Moments of Student distributions}

Due to the symmetry under reflection all the odd moments vanish. For
the second moments we have, provided that $\nu>2$,\[
E(x_{i},x_{j})=\frac{\Lambda_{ij}}{\nu-2}\,.\]
The moments of order $2n$ exist provided that $\nu>2n$ ; as happens
for Gaussian distributions, they can be expressed in term of the second
moments,\begin{eqnarray*}
E(x_{j_{1}},x_{j_{2}},\dots,x_{j_{2n}}) & = & \frac{\Gamma(\frac{\nu}{2}-n)}{2^{n}\Gamma(\frac{\nu}{2})}\prod_{\textrm{all the pairings}}\Lambda_{j_{i_{1}}j_{i_{2}}}\cdots\Lambda_{j_{i_{2n-1}}j_{i_{2n}}}.\end{eqnarray*}

In the univariate case these formulas reduce to $E(x^{2})=\frac{\lambda^{2}}{\nu-2}$
and \[
E(x^{2n})=\frac{(2n-1)!!\Gamma(\frac{\nu}{2}-n)}{2^{n}\Gamma(\frac{\nu}{2})}\lambda^{2n}.\]
The kurtosis is then $\kappa=3\frac{\nu-2}{\nu-4}$, provided that
$\nu>4$.

\subsection*{Simulation of multivariate Student distributions}

The simulation is a standard application of the technique used in
the case of rotational invariance. From \[
g_{n}(\mathbf{x})d^{n}\mathbf{x}=\frac{\Gamma(\frac{\nu}{2}+\frac{n}{2})}{\pi^{n/2}\Gamma(\frac{\nu}{2})}r^{n-1}(1+r^{2})^{\frac{1}{1-q}}d^{n-1}\Omega dr,\]
 with $r\geq0$, we see that the density of the angular variables
is uniform, while setting $y=\frac{r^{2}}{1+r^{2}}$, with $1>y\geq0$
and $r=\sqrt{y/(1-y)}$, the density of $y$ is given by\[
\frac{1}{B(\frac{n}{2},\frac{\nu}{2})}y^{\frac{n}{2}-1}(1-y)^{\frac{\nu}{2}-1}dy,\]
i.e. by the beta distribution with parameters $\frac{n}{2}$ and $\frac{\nu}{2}$.
Eventually we can simulate the multivariate $n$ dimensional distribution
by
\begin{enumerate}
\item Simulating $y$ according to $B_{x}(\frac{n}{2},\frac{\nu}{2})$ and
setting $r=\sqrt{\frac{y}{1-y}}$.
\item Simulating $n$ i.i.d. Gaussian variables $u_{i}$ and settings $\mathbf{n}=(u_{1},\dots,u_{n})/\sqrt{u_{1}^{2}+\cdots+u_{n}^{2}}$.
\item Returning $x\mathbf{n}$.
\end{enumerate}
The more general case \eqref{eq:gmqg} is simulated using the same
algorithm and then returning $\mathbf{C}\mathbf{x}$, where $\mathbf{\Lambda}^{-1}=(\mathbf{C}^{T})^{-1}\mathbf{C}^{-1}$,
i.e. $\mathbf{\Lambda}=\mathbf{C}\mathbf{C}^{T}$.

\subsection*{Characteristic function of symmetric generalized hyperbolic distributions}

We start from the expression\[
f_{n}(\mathbf{x})=\frac{\alpha^{\frac{n}{2}}}{(2\pi)^{\frac{n}{2}}K_{\frac{\nu}{2}}(\alpha)}\frac{K_{\frac{\nu}{2}+\frac{n}{2}}(\alpha\sqrt{1+r^{2}})}{(1+r^{2})^{\frac{\nu}{4}+\frac{n}{4}}},\]
with $r=\sqrt{\sum_{i=1}^{n}x_{i}^{2}}$; the general case is obtained
simply with an affine transformation $\mathbf{x}\rightarrow\mathbf{\mu}+\delta\mathbf{R}\mathbf{x}$,
with $\mathbf{\mu}\in\mathbb{R}^{n}$, $\delta\geq0$ a scale parameter,
and $\mathbf{R}$ an orthogonal transformation in $\mathbb{R}^{n}$.
The central expression we need is an integral of the Sonine-Gegenbauer
type, cf.~identity 7.14.(46) of \citet{erdely2}:\begin{multline*}
\int_{0}^{\infty}dt\, J_{\mu}(bt)K_{\nu}(a\sqrt{t^{2}+z^{2}})(t^{2}+z^{2})^{-\frac{\nu}{2}}t^{\mu+1}\\
=b^{\mu}a^{-\nu}z^{\mu-\nu+1}(a^{2}+b^{2})^{\frac{\nu}{2}-\frac{\mu}{2}-\frac{1}{2}}K_{\nu-\mu-1}(z\sqrt{a^{2}+b^{2}})\\
\Re(\mu)>-1,\;\Re(z)>0.\end{multline*}

For $n=1$, considering that $J_{-\frac{1}{2}}(x)=\sqrt{\frac{2}{\pi x}}\cos(x)$,
we obtain

\begin{multline*}
\tilde{f}_{1}(k_{1})=\int_{-\infty}^{+\infty}dx_{1}\, e^{ik_{1}x_{1}}f_{1}(x_{1})=\frac{2\alpha^{\frac{1}{2}}}{(2\pi)^{\frac{1}{2}}K_{\frac{\nu}{2}}(\alpha)}\int_{0}^{+\infty}dx_{1}\,\frac{K_{\frac{\nu}{2}+\frac{1}{2}}(\alpha\sqrt{1+x_{1}^{2}})}{(1+x_{1}^{2})^{\frac{\nu}{4}+\frac{1}{4}}}\cos(k_{1}x_{1})\\
=\frac{\alpha^{\frac{1}{2}}k_{1}^{\frac{1}{2}}}{K_{\frac{\nu}{2}}(\alpha)}\int_{0}^{+\infty}dx_{1}J_{-\frac{1}{2}}(k_{1}x_{1})\, K_{\frac{\nu}{2}+\frac{1}{2}}(\alpha\sqrt{1+x_{1}^{2}})(1+x_{1}^{2})^{-\frac{\nu}{4}-\frac{1}{4}}x_{1}^{\frac{1}{2}}\\
=\frac{K_{\frac{\nu}{2}}(\sqrt{\alpha^{2}+k_{1}^{2}})}{K_{\frac{\nu}{2}}(\alpha)}\frac{(\alpha^{2}+k_{1}^{2})^{\frac{\nu}{4}}}{\alpha^{\frac{\nu}{2}}}.\end{multline*}
For alternative derivations in the univariate case see \citet{Hu95cfstd}
and the references therein.

In our setting the computation is exactly the same for general $n$,
with $k=\sqrt{\sum_{i=1}^{n}k_{i}^{2}}$, $d^{n-2}\Omega$ the surface
element of the sphere $S^{n-2}$, $\phi$ the angle between $\mathbf{k}$
and $\mathbf{x}$, using identity \eqref{eq:bess2}\begin{multline*}
\tilde{f}_{n}(\mathbf{k})=\int_{\mathbb{R}^{n}}d^{n}\mathbf{x}\, e^{i\mathbf{k}\cdot\mathbf{x}}f_{n}(\mathbf{x})\\
=\frac{\alpha^{\frac{n}{2}}}{(2\pi)^{\frac{n}{2}}K_{\frac{\nu}{2}}(\alpha)}\int d^{n-2}\Omega\int_{0}^{+\infty}dr\, r^{n-1}\int_{0}^{\pi}d\phi\,\sin^{n-2}(\phi)e^{ikr\cos\phi}\frac{K_{\frac{\nu}{2}+\frac{n}{2}}(\alpha\sqrt{1+r^{2}})}{(1+r^{2})^{\frac{\nu}{4}+\frac{n}{4}}}\\
=\frac{k^{1-\frac{n}{2}}\alpha^{\frac{n}{2}}}{K_{\frac{\nu}{2}}(\alpha)}\int_{0}^{+\infty}dr\, J_{\frac{n}{2}-1}(kr)K_{\frac{\nu}{2}+\frac{n}{2}}(\alpha\sqrt{1+r^{2}})(1+r^{2})^{-\frac{\nu}{4}-\frac{n}{4}}r^{\frac{n}{2}}\\
=\frac{K_{\frac{\nu}{2}}(\sqrt{\alpha^{2}+k^{2}})}{K_{\frac{\nu}{2}}(\alpha)}\frac{(\alpha^{2}+k^{2})^{\frac{\nu}{4}}}{\alpha^{\frac{\nu}{2}}}.\end{multline*}

Hence the eventual result $\tilde{f}_{n}(\mathbf{k})=\tilde{f}_{1}(k)$.

\bibliographystyle{plainnat}
\bibliography{econophysics}

\end{document}